\documentclass{vgtc}                          

\ifpdf
  \pdfoutput=1\relax                   
  \usepackage{graphicx}                
  \DeclareGraphicsExtensions{.pdf,.png,.jpg,.jpeg} 
\else
  \usepackage{graphicx}                
  \DeclareGraphicsExtensions{.eps}     
\fi%

\graphicspath{{figures/}{pictures/}{images/}{./}} 

\usepackage{microtype}                 
\PassOptionsToPackage{warn}{textcomp}  
\usepackage{textcomp}                  
\usepackage{mathptmx}                  
\usepackage{times}                     
\usepackage{cite}                      
\usepackage{tabu}                      
\usepackage{booktabs}                  

\usepackage{enumitem}

\usepackage{color}
\definecolor{cb_blue}{rgb}{0.22,0.49,0.72}
\definecolor{cb_green}{rgb}{0.3,0.67,0.29}
\definecolor{cb_orange}{rgb}{0.84, 0.45, 0.06}
\definecolor{cb_red}{rgb}{1.0, 0.0, 0.0}
\definecolor{cb_periwinkle}{rgb}{0.4, 0.2, 0.8}
\definecolor{black}{rgb}{0,0,0}

\usepackage{xurl} 
\usepackage{hyperref}

\onlineid{0}

\vgtccategory{Research}

\vgtcinsertpkg

\title{Data Bricks Space Mission: Teaching Kids about Data with Physicalization}

\author{Lorenzo Ambrosini\thanks{e-mail: ambrosini.lorenz@gmail.com}\\ %
     \scriptsize Link\"{o}ping University %
\and Miriah Meyer\thanks{e-mail: miriah.meyer@liu.se}\\ %
     \scriptsize Link\"{o}ping University}

\teaser{
  \centering
  \vspace{-0.2cm}
   \includegraphics[width=0.95\linewidth]{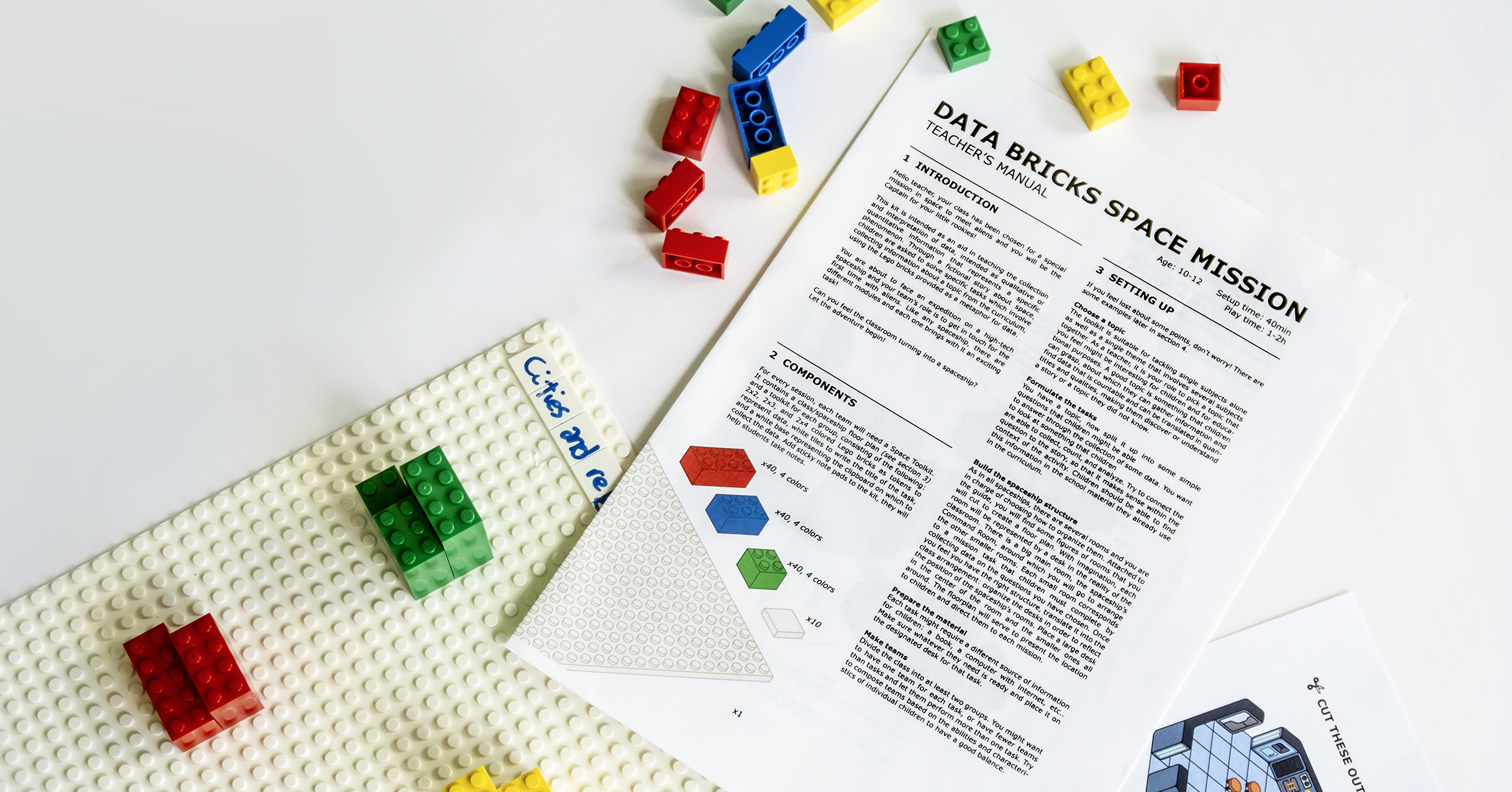}
   \vspace{-0.2cm}
  \caption{The Data Bricks Space Mission is a role-playing physicalization activity for kids aged 10-12 years to engage them with the abstract concept of data. The activity's toolkit includes Lego bricks to represent data and build a visualization and the teachers' manual to guide students through the activity.}
  \label{fig:teaser}
}

\abstract{The Data Bricks Space Mission is a prototype activity based on data physicalization for teaching kids about data.
The design of the activity is based on a literature review and interviews with elementary school teachers, and targets kids aged 10-12. Using Lego bricks and a fictional space adventure story, teachers can use the Data Bricks Space Mission activity to empower kids to produce data, communicate their findings, and gain a better understanding of the relationship between data and the world around them.
\\

\noindent \textbf{key words:} data physicalization, data literacy, children\\%
} 

\begin{document}



\maketitle

\section{Introduction} 







Understanding how we use, consume, and produce data is increasingly important as our work and lives become more intertwined with data. Being able work with data is a vital component of many professional fields, with data skills increasingly being important for success in many jobs. And awareness of the data traces we leave in our use of digital tools has broad implications for our future privacy, security, and control of personal data. The concept of data, however, is abstract and ill-defined, making it difficult for many people to fully understand its implications.

In this research we tackle the challenge of teaching kids about data: what it is, how it is produced, and how we can visualize it to communicate with others. We set out to design an activity for teachers to empower kids to produce data, and to acquire hands on learning about the connection between data and the world around them. The activity builds on data physicalization to engage younger students, and encourages them to make direct links between questions about their lives and how to answer those questions through the production of data.


More specifically, the contribution of this paper is the Data Bricks Space Mission, a prototype activity for teachers to use in the classroom to teach kids about data, shown in Figure \ref{fig:teaser}. The design of the activity is grounded in a literature review of data physicalization and activities that engage kids with data, as well as interviews with elementary school teachers. The activity targets kids aged 10-12, is based on a role-playing story about a space mission to meet aliens, and uses Legos as a data physicalization material. We describe our formative work that inspired the activity in Sections \ref{sec:related_work} and \ref{sec:methods}, the design requirements in Section \ref{sec:requirements}, provide a detailed description of the activity goals and steps in Sections \ref{sec:activity}, and include our interview and activity materials in supplemental materials. 



\section{Related Work}
\label{sec:related_work}


 
The Data Bricks Space Mission activity uses data physicalization to make the abstract concept of data more tangible, legible, and comprehensible. Jansen et al.~\cite{jansen_2015} define \textit{data physicalization} as a \textit{``physical artifact whose geometry or material properties encode data''}. Also referred to as constructive visualization, this approach to creating and visualizing data is characterized by  Huron et al.~\cite{huron_2014} through the following concepts:
 \begin{itemize}[nosep]
     \item \textbf{tokens:} the discrete basic units representing data, 
     with different features such as color, shape, volume, material, etc.
     \item \textbf{token grammar:} the relationship of a set of different tokens and their attributes to aspects of the data
     \item \textbf{environment:} the space in which tokens are placed and the structural criteria for how they can be combined
     \item \textbf{assembly model:} the rules of how the visual representation is built and disassembled. 
 \end{itemize}

Data physicalization employed in research activities and workshops shows that this approach allows people to create and interpret data representations, educate through playful activities, and foster reflection and discussion on a variety of domain subjects. \textit{Viskit}~\cite{huron_2016} is a toolkit composed of materials and a process for conducting data physicalization workshops. Designed for adults learners, the Viskit workshop is an approach to teaching people about visualization outside of a typical classroom setting. In another example, 
Thudt et al.~\cite{thudt_2018} conducted a qualitative study of people who, for 
2-4 weeks, collected personally relevant data using physical tokens of their own choice. Their results show that constructing physical visualizations can be personal, varied, and enmeshed in the everyday lives of participants, and can encourage a broad range of personal reflections. 
A study by Huron et al.~\cite{huron_2017} engaged HCI and design professionals with data physicalization in workshop settings to help participants ideate, prototype, and design with data. Their results indicate that physicalizations make the abstract concepts of data more tangible and graspable. 
Taken together, these studies provide evidence for the effectiveness of data physicalization in helping people collect, understand, and reflect on data; but, these results are limited to adults. The physicalization game \textit{Datablokken}~\cite{verhaert_2021}, on the other hand, is designed specifically for kids, and teaches them how to think critically about their personal data within social media. 
Like Datablokken, the Data Bricks Space Mission is a physicalization activity specifically designed to engage kids, but with the aim of teaching them more broadly about the abstract concept of data.

Other non-physicalization projects seek to teach kids more generally about data and how to use it to answer questions, but most often as part of mathematics or statistics curriculum using paper-based or digital tools. \textit{Data4Kids}~\cite{schwabish_2022} aims to improve children's data literacy through 5 different data stories, each consisting of starter kits at 3 levels for different grades. Each digital kit includes a dataset, a data dictionary with definitions, and slides with suggested topics for data exploration and visualization. 
Hutchison et al.\cite{hutchison_2000} present another project that had kids ask, gather, and answer their own data-driven questions using paper-based methods. Their experiences within the project highlights the difficulty of picking a suitable data-driven question, with teacher suggested topics as unappealing and the kids' suggested topics as not well suited to data collection. They also note that having a facilitator to guide kids through the project is important for success.
In another study, Stornaiuolo~\cite{stornaiuolo_2020} worked with teenagers to reframe their role from passive data producers and consumers into data agents who can curate, gather, and visualize their own personal data stories. In their study students collected data using Google Sheets and crafted visualizations on t-shirts. Their results indicate that data literacy education can enable youths to think critically about how they use data and recognize their own data rights. 
The Data Bricks Space Mission extends these results by using data physicalization as another approach to teach young kids about data and empower them to be their own data producers.

\section{Methods}
\label{sec:methods}



This project was completed as part of the first author's Masters thesis~\cite{ambrosini_2022}. We took a research through design (RtD) approach throughout the project, generating new knowledge through the development of a practical, feasible solution to a problem~\cite{stappers_2017}. RtD is an approach where designers explore potential futures and express the knowledge gained about such possibilities through designed artifacts~\cite{zimmerman2007}. More specifically, we made use of the \textit{double diamond model}~\cite{designcouncil_2019} for our design process in order to get from our initial research questions to an effective solution. This model consist of a series of diverging and converging design phases: discover, define, develop and deliver.


At the start of the project we discussed our personal interests in data visualization, physicalization, and empowering kids. This led to an initial literature review into data physicalization where we learned that most previous research has focused on adults.
We then dug into a second round of literature into activities for teaching kids about data, and found only a limited number of articles and approaches.

Next we conducted several interviews with local primary school teachers.
We iteratively developed an interview guide based on our literature review and own interests within the project, which included several printed materials to explore a range of possible activity designs during the interviews. These hypothetical, example activities considered a number of design dimensions: materials, environment, final outcome, participants, topic choice, purpose, guidance, and freedom. We piloted the interview on a former teacher 
and refined the guide based on feedback. Finally, we conducted interviews with 1 former and 2 current school teachers who teach grades 4-6 in a public primary school in Norrk\"{o}ping, Sweden. We include our interview guide and materials in supplemental materials.



\begin{figure}[h]
    \centering
    \includegraphics[width=\linewidth]{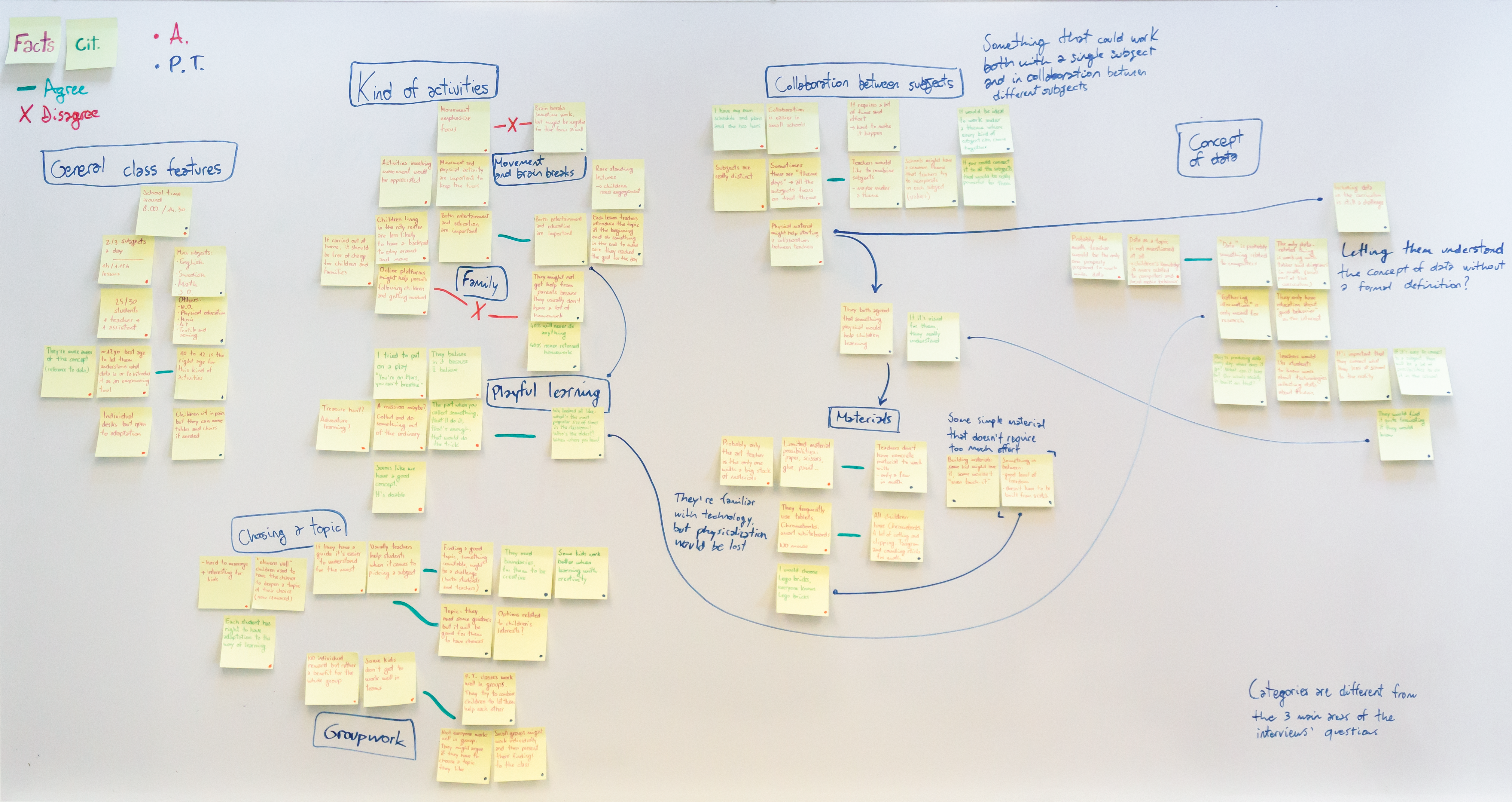}
    \vspace{-0.4cm}
    \caption{The first author created an affinity diagram to organize our insights from interviews with primary school teachers. The organization elucidated several themes that we used as design requirements for our data physicalization activity.}
    \label{fig:affinity-diagram}
    \vspace{-0.4cm}
\end{figure}

The first-author transcribed and summarized the interviews. He then combined the  interview summaries and insights from the literature review into an affinity diagram, shown in Figure \ref{fig:affinity-diagram}, to develop a general set of design requirements for the project. Using these design requirements we iteratively developed the Data Bricks Space Mission activity as a design prototype.


\section{Design Motivation and Requirements}
\label{sec:requirements}



This project is aimed at an audience of children between the ages of 10 and 12. Our interviews with teachers highlighted their excitement to engage with students in this age range about data, confirmed this age range to be capable of grasping the abstract concept of data, and revealed the lack of existing curriculum to teach students about this topic in an engaging and empowering way.
Furthermore,  our interviews revealed that:
\begin{itemize}[nosep]
    \item active experience with physical materials works better than lecturing, but time and access to resources in the classroom are difficult barriers to overcome;
    \item students in this age range need active guidance from teachers during activities;
    \item successful activities are flexible for individual children’s different needs and skills;
    \item group work is welcomed in the classroom;
    \item engaging parents in after-school homework activities would not be possible in all families;
    \item and the topic of data is almost totally absent in the Swedish curriculum of grades 4-6 except to teach kids about the dangers of exposing personal data on social media.
\end{itemize}

Based on these insights, we concluded that the activity should take place at school with the teacher acting as a guide, and that the materials must convey qualities such as simplicity, expressiveness, and dynamism. Additionally, the activity would need precise guidelines and rules so kids (and teachers) do not get lost or confused about how to proceed on their own. The ideal situation is that the activity could be integrated as much as possible into any lessons or subjects, complementing the current curriculum with data-based perspectives. The interviews highlighted opportunities to engage kids through fun, creative, and physical activities that fall outside of standard classroom activities.

\section{Data Bricks Space Mission}
\label{sec:activity}



The Data Bricks Space Mission is a role-playing activity that engages kids through a story about meeting with aliens and communicating knowledge about Earth. In the activity, students divide into teams and collect important information about their language, environment, country, and more. They use Lego bricks to tokenize the information they gather and to communicate their findings back to the rest of the class. The goal of the activity is to support data literacy in young students through engaging them in the production data. 
The activity introduces kids to the concept of data through physicalization, aiming to instill an understanding about the potential opportunities to create and use data in everyday life. 

The activity is designed to help educators teach students how to gather and understand data that represents a specific phenomenon. It is a simplified form of a role-playing game in which the teacher acts as a facilitator and Lego bricks serve as both props for storytelling and a tangible representation of data. Teachers guide the students through specific tasks to collect data on a chosen topic within the curriculum. The game-like nature of the activity and lack of rigid rules make it intentionally open-ended and flexible to accommodate varying learning approaches and student needs. This flexibility provides instructors with a range of opportunities for adaptation to various classroom circumstances, topics, and goals.

Within the activity there is no formal definition for data. Instead, students learn about data implicitly through data collection tasks and group presentations. 
Through a final discussion, the objective is to help children comprehend that data could relate to nearly any school subject. A more general educational objective of the activity is to also help kids realize that what they study in school is not knowledge for knowledge's sake, but rather knowledge of the actual world around them. 


\begin{figure*}[t]
    \centering
    \includegraphics[width=\linewidth]{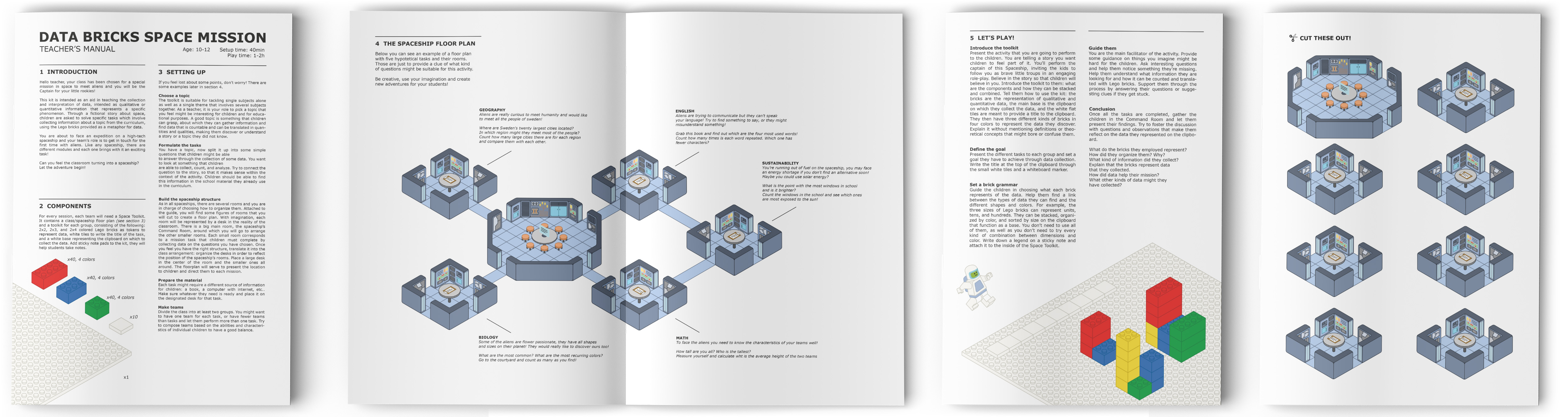}
    \vspace{-0.6cm}
    \caption{The teacher's manual is divided in three main parts and a separate sheet with the cut-outs. The first page presents what the activity is about and how it is prepared. The middle pages provide an example of the spaceship's floor plan with example data-collection questions for each room and group of students. The last page explains how to conduct and conclude the activity, including suggested points for classroom discussions that relate the activity to the concept of data. A PDF of the manual is included supplemental materials.}
    \label{fig:space-mission-guide}
    \vspace{-0.4cm}
\end{figure*}




\subsection{The Activity and Toolkit}

The Data Bricks Space Mission activity is designed for educators who work with students between the ages of 10 and 12. The activity takes place in the classroom and is designed to be included within an existing curriculum for any subject. 
Students are asked to join a special mission aboard a spacecraft under the leadership of their Captain -- the teacher -- to start a conversation with aliens. Each aspect of the activity contributes to the development of the fictitious narrative, in which kids are given a variety of tasks to do aboard a spacecraft which include gathering data in order to provide an answer to a range of questions. 
To answer their assigned questions, students are expected to do research and record their findings using Lego bricks. 
The student-led research can be conducted both in the class and outside, if required, depending on the task. It is suitable for both a single specific subject as well as a planned interdisciplinary activity, in the case that teachers desire to collaborate on a common school theme such as sustainability. 

The Data Bricks Space Mission toolkit offers the tangible supplies needed to make the story and activity possible, with the teacher's imagination filling in the gaps.
The toolkit includes a teacher's instruction manual -- shown in Figure \ref{fig:space-mission-guide} and included in the supplemental materials -- describing the contents of the kit and the recommended procedure to follow while planning for and carrying out the activity. 
The manual encourages the teacher to rearrange the classroom into the fictionalized spacecraft. Each desk can represent a room on the spaceship where a specific task is performed. A sample classroom-spaceship floor plan is supplied in the instruction manual, along with example tasks and questions for the activity. Using the paper cutouts of the rooms included in the manual, the instructor creates a floor plan that matches the tasks they intend to do. 
Finally, the manual illustrates how the teacher can help students think about bricks as data tokens by introducing a grammar that defines the metaphors between data and its representation.

In addition to the teacher instruction manual, the toolkit is composed of cardboard suitcases that each group of students takes with them on their expedition. Each suitcase includes a set of Lego bricks composed as follows: 20 2x2 bricks, in 4 colors; 20 2x3 bricks, in 4 colors; 20 2x4 bricks, in 4 colors; 1 32x32 white base; and 10 2x2 flat white tiles. Lego bricks were chosen for the toolkit as they offer a simple yet adaptable approach to data physicalization. This combination of Legos provides 3 alternative dimensions to work with: 4 distinct colors to communicate different qualities, 3 different brick sizes, and the number of bricks chosen. The Lego base serves as a clipboard, on which children can annotate the data and thereby construct the physicalization. The clipboard 
creates an environment with three more parameters, similar to a three-dimensional Euclidean space, with width, length, and height. The white smooth surface of the flat tiles can be placed at the top of the clipboard to allow the students to provide a title for their physicalization using a whiteboard marker. The toolkit additionally contains sticky notes and a whiteboard marker. This toolkit design is intended for reuse.

\subsection{Setup}

The setup time for the activity by the teacher is expected to take about 40 minutes, with the activity taking between 1-2 hours.
Guided by the examples in the manual, the teacher selects topics for the various spaceship rooms and articulates tasks for the students to complete. The task should consist of a simple question that kids can answer by gathering information. 
Teachers are encouraged to pick a range of topics that  work together in the broader story of the activity.
The teacher must also make available the informational resources that the students will use to gather information, such as textbooks, the internet, the school library, lecture notes, or access to outdoors. 
Example tasks and questions included in the manual are:
\begin{itemize}[nosep]
    \item Geography: Aliens are really curious to meet humanity and would like to meet all the people of Sweden! Where are Sweden's twenty largest cities located? In which region might they meet most of the people? Count how many large cities there are for each region and compare them with each other.
    \item Biology: Some of the aliens are passionate about flowers. They have all shapes and sizes on their planet, and they would like to discover ours too! What are the most common? What are the most recurring colors? Go to the courtyard and count as many as you find!
    \item Language: Aliens are trying to communicate but they can’t speak your language! What words should you teach them?
    Grab a book and find the four most used words. Count how many times each word is repeated. Which word has the fewest characters?
    \item Sustainability: You’re running out of fuel on the spaceship and you need to find an alternative source! Maybe you could use solar energy? What is the room with the most windows in school and is it brighter? Count the windows around school and note which ones are most exposed to the sun!
    \item Math: To face the aliens you need to describe your team well! How tall are you all? Who is the tallest? Measure yourselves and calculate the average height of your team.
\end{itemize}
These examples are provided in the manual as inspiration for teachers to use in creating their own questions and tasks suitable for use in their classes.

Next, the teacher prepares the physical space where the activity will take place. The teacher creates the spaceship's floor layout using the available cutouts, assigning each spaceship room a task, and re-arranges the classroom accordingly.
Each desk represents a different room in the fictional world, with the Command Room being represented by a bigger desk in the middle. Each spaceship room receives a toolkit suitcase with the data physicalization supplies.

The teacher divides the students into groups as the final step of preparation. To create groups that are equal and balanced, the division could be done in accordance with the children's preferences and skills. Whether or if the instructor finds it to be an engaging and formative exercise for the kids will determine whether or not this entire initial phase is carried out with their participation.

\subsection{Play}

When everything is set, the teacher creates the role-playing story for the students. As the spaceship's Captain, the teacher invites the students to join them as brave young troops. The educator should go through the toolkit's components and demonstrate how to stack and combine the data bricks. 
The educator then assists the students in defining the purpose of each groups' task on their clipboards and the grammar of their bricks by making a connection between the various shapes and colors and the sorts of information that may be discovered. To facilitate the process, the grammar can be noted with sticky notes attached to the kit box, so that children always have a reference in front of them.

The students are then instructed to start collecting their information. At this point the teacher's role is that of a facilitator, posing questions and identifying areas where students may be having difficulty or failing to notice things they could measure or count. After each group completes their data collection tasks, the teacher collects the groups in the Command Room and asks them to share their results with the other students. The teacher can encourage conversation by posing inquiries and making observations that prompt the students to consider the information they portrayed on the clipboard. The mission is successful and the aliens forge a fruitful partnership with Earth if the data physicalizations produced by the groups accurately depict the student's observations. 
Finally, the instructor has the option of engaging the students in a discussion about how the information they gathered could be collected or represented differently, and to discuss the relationship of their data bricks to the concept of data more generally. To do this, the manual suggests posing questions like: \textit{``What do the bricks represent?''} or \textit{``How did data help your mission?''}.




\section{Next steps}

The Data Bricks Space Mission is an initial prototype, designed from the insights we acquired through a literature review and interviews with several teachers. The next step for this research is to get feedback on the design.
Initial feedback could come from the teachers who took part in our interviews, with the scope then expanding to other educators. Expanding the network of professionals engaged could make it more likely that new ideas and improvements will emerge. Interviews demonstrated that literacy and education are dependent on outside circumstances that call for adaptability and flexibility. Various concepts would thus undoubtedly result in a deeper and more flexible solution for other scenarios.
Additional design iterations and considerations could also be explored for the physical components of the toolkit. Depending on feedback from educators, it is possible that Lego bricks might be replaced or enhanced with different materials. 
We anticipate that feedback from educators would provide insights on how to improve all aspects of the toolkit.

In recent years, remote teaching has become an increasingly present reality, supported by digital tools, as an alternative to classic classroom lessons. In this context, it would be interesting to explore what a virtual version of this activity could be. As not all kids might have Lego bricks in their homes, it would be necessary to find other tokens to replace the Lego bricks. Group work would also need to be reconsidered and redesigned.
Despite this, the qualities of entertainment and storytelling that do not depend on the medium through which the experience is addressed could remain unchanged. 

Finally, the activity needs to be tested and refined based on a deployment in classrooms. A testing phase will clarify how the activity can be useful to education, as well its strengths and flaws. These tests further look at the effectiveness of data physicalization for engaging kids with data and visualization.

\section{Conclusions}
In this paper we present the Data Bricks Space Mission, an activity and toolkit to support data literacy among children through physicalization. Combining the knowledge gained from literature and interviews with educational professionals, we present the prototype of an activity that is intended for use in primary schools. The activity represents a step towards promoting more discussions about how to build a fruitful data education experience among youths.


\acknowledgments{
The authors wish to thank the educators who gave their time and input for this project, as well as the anonymous reviewers for their feedback.
This work is partially funded by the Wallenberg AI, Autonomous Systems and Software Program (WASP) funded by the Knut and Alice Wallenberg Foundation.}

\bibliographystyle{abbrv-doi}

\bibliography{data-bricks}
\end{document}